\documentclass[original]{elsarticle}
\usepackage{lineno,hyperref}
\usepackage{amsmath,amssymb,amsfonts}
\usepackage{algorithmic}
\usepackage{graphicx}
\usepackage{textcomp}
\usepackage{caption,setspace}
\journal{Under Review}

\begin{document}
	
	\begin{frontmatter}
		
		\title{Minus HELLO: HELLO Devoid Protocols for Energy Preservation in Mobile Ad-hoc Networks}
	\author[add1]{Anuradha Banerjee}
	\author[add2]{Abu Sufian \corref{cor1}}
	\ead{sufian@ugb.ac.in}
	\author[add3]{Paramartha Dutta}
	\author[add4]{M M Hafizur Rahman}
	\cortext[cor1]{Corresponding author}
	\address[add1]{ Department of Computer Application,  Kalyani Govt. Engineering College, Kalyani, India}
	\address[add2]{ Department of Computer Science, University of Gour Banga, Malda, India}
	\address[add3]{Department of Computer and System Sciences, Visva-Bharati, Santiniketan,  India }
	\address[add4]{Dept. of Computer Networks and Communications, CCSIT, King Faisal University, Al Ahsa 31982, Saudi Arabia.}
		
		
		
		
\begin{abstract}
In mobile ad-hoc networks, nodes have to transmit HELLO or Route Maintenance messages at regular intervals, and all nodes residing within its radio range, reply with an acknowledgment message informing their node identifier, current location, and radio-range. Regular transmitting these messages consume a significant amount of battery power in nodes, especially when the set of down-link neighbors does not change over time and the radio-range of the sender node is large. The present article focuses on this aspect and tries to eliminate the number of HELLO messages in existing state-of-art protocols. Also, it shortens radio-ranges of nodes whenever possible. Simulation results show that the average lifetime of nodes greatly increases in proposed Minus HELLO devoid routing protocols along with a great increase in network throughput. Also, the required number of route re-discovery reduces. 

\end{abstract}
		
		
\end{frontmatter}
	

\section{INTRODUCTION}

A mobile ad-hoc network or simply MANET is an infrastructure-less network consisting of only some mobile nodes that move freely in any direction \cite{chl03}, \cite{conti14}. These networks can be deployed in emergency situations like war, natural disasters, etc \cite{helen14}. In addition, MANET is useful for communication in the Internet of Things(IoT) \cite{alam2020internet, zikria2018opportunistic}. Battery-powered nodes act as endpoints or routers to selflessly forward packets in a multi-hop environment \cite{vazi14, Sufian2019}.  Therefore, energy efficiency in every node is crucial to preserve the battery power of nodes and increase their lifetime \cite{Corson99, benerjee17, khamayseh2018ensuring, thiyagarajan2020power}. In this perspective, this article presents the following contributions.  

\subsection{Contributions of this Proposed Scheme}
i) The present article proposes a novel idea of reforming routing protocols so that they can work without or very little HELLO messages. \textbf{M}inus HELLO (-HELLO) emphasizes that being informed about a downlink neighborhood is absolutely unnecessary until and unless a node participates in a communication session as an endpoint or router. For simplicity of the representation, we shall refer to Minus HELLO as -HELLO in the rest of the article.\\
ii)	A general framework of -HELLO version of state-of-the-art protocols is presented with mathematical illustration as case studies.\\
iii) Message formats of state-of-art protocols have been re-designed to contain certain additional attributes to overcome the absence of regular HELLO message exchanging.\\
iv)	It has been shown in our article that hidden and exposed terminal problems can be resolved without HELLO messages exchanging. Therefore, the robustness of the network does not suffer.\\
v) Detailed simulation results emphasize that the communication protocols without HELLO messages, save a lot of energy with a significant increase in network throughput.
\subsection{Organization of the article}
The organization of the rest of the article is as follows. Section \ref{lts} deals with a brief description of various routing protocols in MANETs, including proactive, reactive, and energy-efficient. General Methodology of -HELLO \ref{gi}. Section \ref{cs} explains -HELLO devoid protocols as case studies, such as -HELLO versions of the protocols AODV \cite{perkins99}, MMBCR \cite{toh01}, MRPC \cite{misra02}, MTPR \cite{natarajan09} and MFR \cite{hou86}. Here we have omitted proactive routing protocols because they are not suitable for large networks that are when the number of nodes is high. AODV and MFR are two state-of-the-art representatives of reactive routing protocols, whereas MMBCR, MTPR, and MRPC are three popular representatives of energy-efficient protocols. Section \ref{sm} presents the simulation results while section \ref{con} concludes the paper.

\section{EXISTING STATE-OF-THE-ART ROUTING PROTOCOLS}
\label{lts}
The literature of MANETs is rich in proactive, reactive, and energy-efficient protocols \cite{abolhasan04}. Destination-sequenced Distance Vector (DSDV) \cite{perkins94}, Cluster-based Gateway Switch Routing (CGSR), Global State Routing (GSR) \cite{chen98}, The wireless routing protocol(WRP) \cite{murthy96}, Fisheye state routing (FSR)\cite{853066} etc. are state-of-the-art proactive protocols. These instruct the nodes to store route information to every other node in the network. Hence, a regular update of routing tables is required, consuming huge battery power as well as bandwidth.
 
Among reactive routing protocols, ad hoc on-demand distance vector (AODV) \cite{perkins99}, dynamic source routing (DSR) \cite{johnson01}, flow-oriented routing protocol (FORP) \cite{su98}, The temporally ordered routing algorithm (TORA)\cite{park97}, The operation of location aided routing (LAR)\cite{ko98}, Most forward with fixed radius or MFR\cite{hou86} etc. have become standard. Here routes are discovered on-demand through a RREQ and route-reply(RREP) cycle. A RREQ packet reaches the destination through multiple paths. Among them, one is elected by the destination according to these routing protocols, and sent to the source through a RREP packet, so that the source node can start sending data packets to the intended destination through the best-chosen path.

For energy conservation schemes, a variety of schemes comes with different energy-saving strategies. Some are based on the concept of adjusting the radio range of senders in each hop\cite{perkins99}.  The maximum residual packet capacity (MRPC)\cite{misra02} selects the path that has the maximum number of packets to transmits. This computation is based on the residual energy of nodes involved in the path. Minimum battery cost routing (MBCR) aims to find a route with a maximum remaining battery capacity. The cost of a node is ($1/residual\_battery\_power$), and the cost of a route is a summation of the costs of all its nodes. The route with the minimum cost is selected for communication.
Min-max battery cost routing (MMBCR)\cite{toh01} assigns a performance index of a route with a minimum of battery powers of all nodes in the route. Among multiple routes through which a packet arrives at the destination, the one with the maximum performance index is chosen for communication.
Minimum Transmission Power Routing (MTPR)\cite{natarajan09} selects the path with minimum transmission power, for transferring data packets. The computation of minimum transmission power is done according to Frii's transmission equation \cite{natarajan09}.
An energy harvesting technique is proposed in \cite{arafa18} where the transmitter changes its location to identify better energy harvesting spots and this harvesting energy is utilized for actual data transmission by the current transmitter. Some schemes proposed sleeping strategies for saving energy of nodes. In \cite{benerjee18_1} exhausting nodes are allowed to go to sleep for a pre-defined time period, after which they wake up and resume communications.

Some routing schemes other than the above three also have an existence in literature. Flow oriented routing protocol or FORP\cite{abolhasan04} is a stable path routing protocol that produces comparatively stable paths compare to earlier protocols.  In FESC \cite{benerjee18} a stable single-hop clustering scheme has proposed. Here more battery-powered but less mobile nodes are elect for the cluster head and all other nodes directly connected to nearest cluster head. In a study \cite{ALI201850}, authors introduce new energy and load-aware routing methods based on DSR protocol. They experimented with their scheme in the NS-2 network simulator and shown significant improvement. In SR-MQMR \cite{mina19}, the authors try to increase stability and energy efficiency through multipath routing. Residual energy and velocity(mobility) based routing protocol proposed in \cite{Sufian2019}. 
Associativity Based Routing (ABR) protocol\cite{toh97} where beacons are exchanged periodically between neighbors. In \cite{ROBINSON2019101896}, the authors have proposed a rebroadcast algorithm based on knowledge of neighbors for minimizing the routing overhead and improve the quality of service in MANET. Their proposed algorithm selects routes based on the minimum amounts of delay and good stability. The location of node play significant roles for energy-saving and that are studies in \cite{AKBULUT2019101945}.
 
\section{GENERAL METHODOLOGY OF -HELLO}
\label{gi}
\subsection{Background and basic idea of the scheme}
In MANET, nodes regularly broadcast HELLO messages within their respective radio-ranges, to gain information about their one-hop downlink neighborhood. All nodes lying within the radio-circle of the sender of the HELLO message, reply with acknowledgment or ACK informing their unique identification number, location, radio-range, etc. HELLO, messages are useful from a communication perspective because they enable a node to be aware of available links, among which one is chosen as per the performance metric of the underlying protocol. Also hidden and exposed terminal problems are tackled in MANETs with the help of HELLO messages.
But this HELLO dependency has a disadvantage too. HELLO and ACK messages are exchanged by each node at regular intervals even when the node is not initiating a communication. This eats up huge energy in nodes and reduces their lifetime \cite{mohseni10}.

Our present article focuses on this particular problem. It aims at redesigning state-of-the-art representatives of routing protocols so that they can work without HELLO messages without losing the robustness of the networks. -HELLO points out the fact that this information is irrelevant until and unless a communication request arrives at the node. In -HELLO, neighborhood information is collected during the broadcasting of RREQ.  Eliminating irrelevant HELLO and ACK messages contribute to saving a huge amount of energy in the network. This will improve the average lifetime of nodes. As a result, link breakages due to node battery exhaustion will be reduced up to a great extent. Therefore, the number of RREQ messages injected into the network will be greatly reduced for -HELLO version protocols.

\subsection{Mathematical Analysis of -HELLO devoid version protocols}
-HELLO devoid protocols particularly try to eliminate HELLO messages from reactive and energy-efficient routing protocols. It focuses on the fact that information about neighbors of a node is typically required during route discovery, that is, at the beginning of a communication session when the route to a specific destination is to be found out. For that purpose, the size of RREQ messages in various protocols increases a bit. The case study in the next section shows how HELLO messages can be eliminated or reduced for performance improvement in various standard routing protocols in MANETs, like AODV, MBCR, MTPR, etc. Below we mathematically demonstrate improvements that can be produced by -HELLO  versions of protocols in respect of lifetime, throughput, delay, etc.
Let us denote by $ln(i)$ the link between two nodes $n_i$ and $n_{i+1}$. The status of each node is either up or down. If a node $n_i$ is operational, then its status will be up; otherwise down. According to the study of the discharge curve of batteries heavily used in MANETs, at least 40\% of total battery power is required to remain in operational condition \cite{benerjee10}. Therefore, if $max\_eng(i)$ and $res\_eng(i,t)$ denote maximum and residual energy of node $n_i$ at time $t$, then $n_i$ will be up provided condition in equation \ref{eqn1} is true. 

\begin{equation}
res\_eng(i,t) > (0.4 \times max\_eng(i)) 
\label{eqn1}
\end{equation}

$RT(n_i, n_{i+1})$ is a random variable which indicates liveliness of $ln(i)$ from the perspective of mobility. It indicates potential communication capability of the link from $n_i$ to $n_{i+1}$. It is 1 if $n_{i+1}$ is in radio-range of $n_i$, otherwise it is 0. Then, probability that a route $ROUTE_{(s,d)}$ from $n_s$ to $n_d$ is live, is denoted by $P(ROUTE_{(s,d)})$, and  $P(n_i)$ denotes probability of node $n_i$ is live. Mathematical expression of this appears in equation \ref{eqn2}.
\begin{equation}
P(ROUTE_{(s,d)}) = \prod_{0}^{m}P(n_i)\prod_{0}^{m}RT(n_i, n_{i+1})
\label{eqn2}
\end{equation}
-HELLO version of protocols cannot improve mobility oriented stability of links, that is, $RT(n_i, n_{i+1})$, but it greatly enhances  $P(n_i)$ for all i s.t. $0 <i < m$, as shown in the following lemmas.\\\\
\textbf{Lemma 1}: -HELLO embedded version protocols greatly enhance the lifetime of nodes.\\ 
\textbf{Proof}: Assume that the minimum, maximum and average values of minimum receive powers of nodes in the network are given by $min\_min\_rcv$, $max\_min\_rcv$ and $avg\_min\_rcv$ respectively.  
Hence average transmission power $avg\_trans(i)$ of $n_i$ to process a call, is formulated in equation \ref{eqn3}.
\begin{equation}
avg\_trans(i) = \frac{avg\_min\_rcv (0+R_i^2)}{ 2C} 
\label{eqn3}
\end{equation}
where $R_i$ is radio-range of $n_i$ and $C$ is a constant depending on medium. \\
Let $L(i)$ be the lifetime of $n_i$ in HELLO version protocols, $y$ be the number of HELLO messages transmitted by each node in the network per unit time, and $broad(i)$ is the unit of energy required to broadcast a message. Total number of HELLO messages transmitted by $n_i$ throughout its lifetime, is given by $(y \times L(i))$ and the corresponding energy required to broadcast, is $(y \times L(i) \times broad(i))$ units. Amount of energy units, $n_i$ consumes for processing calls throughout its lifetime, is $(rt(i) \times L(i) \times avg\_trans(i))$ units where $rt(i)$ denotes the total number of message packets is transmitted. Throughout the lifetime of a node $n_i$, it can use only $0.6 \times max\_eng(i)$ amount of energy. Hence it find equation \ref{eqn4}, 
\begin{multline}
L(i) = \frac{(0.6 \times max\_eng(i))}{(y\times broad(i) + rt(i)\times avg\_trans(i))} 
\label{eqn4}
\end{multline}

As far as -HELLO version protocols are concerned, let $L^\prime(i)$ be lifetime of $n_i$, and it can be mathematically formulated in equation \ref{eqn5}.
\begin{equation}
L^\prime(i) = \frac{(0.6 \times max\_eng(i))}{( rt(i) \times avg\_trans(i))}
\label{eqn5} 
\end{equation}

Improvement in lifetime produced by -HELLO version protocols, is calculated by $(L^\prime(i) - L(i))$ and it is given in equation \ref{eqn6}.
\begin{multline}
L^\prime(i) - L(i) = (0.6 \times max\_eng(i))\{ \frac{1}{( rt(i)\times avg\_trans(i))}\\ - \frac{1} {(y broad(i) + rt(i)\times avg\_trans(i))}\}
\label{eqn6}
\end{multline}

i.e. $L^\prime(i) - L(i) > 0 $ \\                                                                                                                            
So, improvement is produced in terms of lifetime.
Without any loss of generality we can assume that, in the route $ROUTE_{(s,d)}$ from $n_s$ to $n_d$, $n_i$ is the node with minimum residual lifetime. Assuming each packet requires $tme$ time duration to reach from source to destination through this path. If $pac$ is the number of packets is to be transferred through this route, then each node in $ROUTE_{(s,d)}$ should be live for at least $(tme \times pac)$ time duration to avoid route-breakage due to battery exhaustion in nodes. If $L^\prime(i) \ge (tme \times pac)$ and 
$L(i) < (tme \times pac)$, then converting a protocol to its -HELLO version will reduce the number of route re-discovery sessions in the network.  A large number of such route request packets will not require to transfer, which substantially reduces node lifetime. Therefore, it is proved that  -HELLO version of protocols greatly enhance the lifetime of nodes.\\\\
\textbf{Lemma 2}: -HELLO version protocols reduce average waiting time of a packet in message queues of nodes.\\
\textbf{Proof}: Let call arrival and departure rates at $n_i$ in -HELLO versions protocols are given by $\lambda^\prime(i)$ and $\mu^\prime(i)$, and for classical protocols$\lambda(i)$ and $\mu(i)$. Then, from the Little's law, average waiting time of a call forwarding request at $n_i$ in -HELLO version of protocols is denoted by $avg\_wait_{(-HELLO)}(i)$ and defined in equation \ref{eqn7}.
\begin{equation}
avg\_wait_{(-HELLO)}(i) =\frac{\lambda^\prime(i)} {\{\mu^\prime(i) \times (\mu^\prime(i) -  \lambda^\prime(i))\}}
\label{eqn7}
\end{equation}
Similarly, average waiting time of a call forwarding request at $n_i$ in the classical  protocols is denoted by $avg\_wait(i)$ and defined in equation \ref{eqn8}. 
\begin{equation}
avg\_wait(i) =\frac{\lambda(i)}{ \{\mu(i) (\mu(i) - \lambda(i))\}} 
\label{eqn8}
\end{equation}
Call arrival rates increase along with increase of route rediscovery session. Hence, $\lambda(i) > \lambda^\prime(i)$. Let
\begin{equation}
\lambda(i)= \lambda^\prime(i) + \Delta \lambda; s.t. \Delta \lambda >0
\end{equation}
But call departure rate remains same, that is, $\mu^\prime(i) = \mu(i)$, because forwarding capacity of nodes do not change.

Then, $avg\_wait(i) - avg\_wait_{(-HELLO)}(i) = \frac{F1(i)}{F2(i)}$

Where $F1(i) = (\lambda^\prime(i) + \Delta \lambda) (\mu^\prime(i) -  \lambda^\prime(i)) - \lambda^\prime(i) (\mu^\prime(i) -   \lambda^\prime(i) + \Delta \lambda)$\\
i.e. $ F1(i) = \Delta \lambda \mu^\prime(i)$

$F2(i) =  \mu^\prime(i) (\mu^\prime(i) -  \lambda^\prime(i) -\Delta \lambda)(\mu^\prime(i) - \lambda^\prime(i)) $                                                            
So, calculation is done by equation \ref{eqn10}
\begin{multline}
	avg\_wait(i) - avg\_wait_{(-HELLO)}(i) = \frac{\Delta \lambda}{ \mu^\prime(i)}\\ -  \lambda^\prime(i) - \Delta \lambda(\mu^\prime(i) -  \lambda^\prime(i)) 
	\label{eqn10}
\end{multline}
Hence, $(avg\_wait(i) - avg\_wait_{(-HELLO)}(i)) > 0$\\ \\
\textbf{Lemma 3}: A node acting according to -HELLO version of protocol, produces higher network throughput than classical protocol.\\
\textbf{Proof}: It has already been mentioned that in the classical versions of protocols, call arrival rates to increase due to an increase in a number of route rediscovery sessions, message contention, and message collision. Hence, $\lambda(i) > \lambda^\prime(i)$. \\
$\lambda(i) > (\lambda^\prime(i) + \Delta \lambda)$

From Little’s law, average number of message forwarding requests in message queue of a node $n_i$ with message queue size $mq(i)$ is denoted by $avg\_req(i)$  and defined in equation \ref{eqn11}. Similarly, the average number of message forwarding requests in the message queue of the same node in -HELLO devoid version protocol is denoted by $avg\_req_{(-HELLO)}(i)$  and defined in equation \ref{eqn12}. 
\begin{equation}
avg\_req(i) = \frac{\lambda^2(i)} {\{\mu(i) (\mu(i) - \lambda(i))\}} 
\label{eqn11}
\end{equation}                                                       \begin{multline}
avg\_req_{(-HELLO)}(i) = \frac{(\lambda^\prime)^2(i)}{\{\mu^\prime(i) (\mu^\prime(i)-\lambda^\prime(i))\}} 
\label{eqn12}
\end{multline}                                     
It already assumed $\mu(i) = \mu^\prime(i)$.
Therefore, equation \ref{eqn13}: 
\begin{equation}
avg\_req(i) - avg\_req_{(-HELLO)}(i) = \frac{F1^\prime(i)}{F2^\prime(i)}
\label{eqn13}
\end{equation}
Where, $F1^\prime(i) = {\lambda^2(\mu(i) -  \lambda^\prime(i)) + \Delta\lambda \lambda^\prime(i)  (2\mu(i) - \lambda^\prime(i))}$ \\                                 
Hence, $F1^\prime(i) > 0$ and $F2^\prime(i) = F2(i)$.
So, 
$avg\_req(i) > avg\_req_{(-HELLO)}(i)$.\\                                                                                                                   
From the point of view of packet loss, following different cases arise where classical and -HELLO version protocols compete. We inspect the cases individually and prove that our protocols perform better.\\
\textbf{Case-1}: Packet loss in  -HELLO version is higher than the classical versions.
The determination of possible conditions for this case is shown below. 

Packet overflow is taking place in both classical and -HELLO versions. Therefore, the average number of message requests in both is higher than the message queue capacity $mq(i)$ of $n_i$.  
Hence, 
\begin{equation}
avg\_req(i) - mq(i) = k1; s.t. k1> 0  
\label{eqn14}
\end{equation}
\begin{equation}
avg\_req_{(-HELLO)}(i) - mq(i) = k2; s.t. k2> 0  
\label{eqn15}
\end{equation}
Approximate packet loss in $n_i$ in classical protocols and -HELLO version are given by $k1$ and $k2$ respectively, such that $k2> k1$, as per assumption in the case.
Subtracting equation \ref{eqn14} from equation \ref{eqn15}, we get equation \ref{eqn16}.
\begin{equation}
 avg\_req_{(-HELLO)}(i) - avg\_req(i) = k2 - k1
 \label{eqn16}
\end{equation}                                                            But here right side  is greater than zero whereas left side is less than zero, which is not possible. Therefore, claim in the statement of case-1 stands false.\\
\textbf{Case-2}: -HELLO version suffers from packet loss while classical is free from packet loss. The situation can be mathematically modeled in equation(17) and in equation \ref{eqn15} as below:
\begin{equation}
avg\_req(i) - mq(i) = -k1;   k1> 0  
\label{eqn17}
\end{equation}

Subtracting equation \ref{eqn15} from equation \ref{eqn17},  we get equation \ref{eqn18},
\begin{equation}
avg\_req(i) - avg\_req_{(-HELLO)}(i)  = -k1-k2 
\label{eqn18}   
\end{equation}
Here also claim in the statement of case-2 stands false as case-1.\\
\textbf{Case-3}: Classical versions suffer from packet loss while -HELLO version is free from packet loss.
The situation can be mathematically modeled in equation \ref{eqn14} and equation \ref{eqn19}.
\begin{equation}
	avg\_req_{(-HELLO)}(i) - mq(i) = -k2;    k2> 0  
	\label{eqn19} 
\end{equation}
Subtracting equation \ref{eqn19} from equation \ref{eqn14} we get,
\begin{equation}
	avg\_req(i) - avg\_req_{(-HELLO)}(i)  = k1+k2    
	\label{eqn20}
\end{equation}
Both sides of the equation \ref{eqn20} are positive, so, case 3 is quite possible.\\
\textbf{Case-4}:The loss of packets in -HELLO version is smaller than the classical versions.
This situation is also formulated as in equations \ref{eqn14} and \ref{eqn15}.
But $k2 < k1$, as per assumption in this case.
With the help of previous equations, it can be seen that both left and right sides of equation \ref{eqn18} are less than zero, which is quite possible for this case. Hence, the claim that appears in the statement of this case is true. \\
\textbf{Case-5}: The loss of packets in -HELLO version is exactly the same as in classical versions. But here, $k1 = k2$ because packet loss of $n_i$ in both classical and -HELLO version are the same. Therefore, the right side of the equation \ref{eqn18} becomes zero while the left side non zero, which is not possible. So, the statement in the current case is false.
\subsection{Hidden terminal detection in -HELLO}
\begin{figure}[!htb]
	\centering
	\includegraphics[width=.6\linewidth]{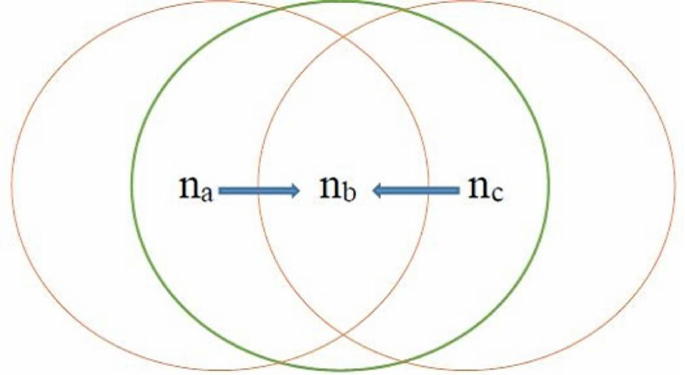}
	\caption{Hidden Terminals}
	\label{Hidden Ter}
\end{figure}
In the figure \ref{Hidden Ter}, $n_a$, $n_b$ and $n_c$ are three different nodes such that the pairs ($n_a, n_b$) and ($n_b, n_c$) can hear each other but $n_a$ and $n_c$ cannot hear each other. Therefore, if $n_a$ and $n_c$ simultaneously send messages to $n_b$, then a signal collision occurs at $n_b$ resulting into loss of messages which is undesirable. This takes place because $n_a$ and $n_c$ are hidden from one another. This is termed as hidden terminal problem.
\subsubsection{Hidden terminals are detected using classical protocols}
Many active detection mechanisms are presented to discover hidden terminals using HELLO messages \cite{li06}. Whenever a node $n_i$ wishes to discover hidden terminals, it unicasts a detection request packet to all of its single-hop neighbors. Those neighbors of $n_i$ unicast probe packets to their respective one-hop neighbors for a time interval mentioned in the detection request of $n_i$. If the waiting time of the detection node expires without receiving an ACK, then the destination of the corresponding detection probe is assumed to be hidden. In this way, a list of hidden terminals is generated. 

The main importance of the HELLO message lies in the fact that detection request and detection probe packets are unicast to one-hop neighbors and if a node $n_i$ needs to know about its one-hop neighbors, it has to rely on HELLO messages that are broadcast at regular intervals within radio-circle of $n_i$.

\subsubsection{Hidden terminals are detected using -HELLO version protocols}
-HELLO versions of protocols follow the similar active mechanism stated in the previous subsection with a simple modification. Here detection requests and probe packets have to be broadcast within the radio-circle of a node so that it reaches all of its 1-hop neighbors. This may require a bit more energy than multiple unicasting of those packets, especially when the number of 1-hop neighbors of the node is small. But unicasting detection request and probe packets are not possible without at least one previous broadcast of HELLO message. Therefore, the overall cost of hidden terminal detection in classical protocols with more than one HELLO message between any two consecutive detection requests or probes is much smaller than the same in protocols based on -HELLO concept. But, if there is exactly one HELLO message between any two consecutive detection requests, the cost of hidden terminal detection with HELLO will be the same as protocols based on -HELLO concept. 

\subsection{Exposed terminal detection in -HELLO}
\begin{figure}[!htb]
	\centering	
	\includegraphics[width=.7\linewidth]{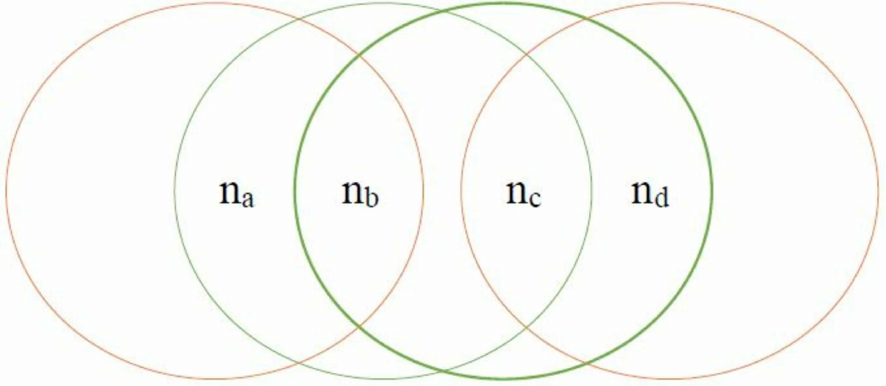}
	\caption{Exposed Terminals}
	\label{Expos ter}
\end{figure}
In the figure \ref{Expos ter}, $n_a$, $n_b$, $n_c$ and $n_d$ are four different nodes such that the pairs ($n_a, n_b$), ($n_b, n_c$), ($n_c, n_d$) can hear each other but the pairs ($n_a, n_c$), ($n_b, n_d$), ($n_a, n_d$) cannot hear each other. Therefore, if $n_b$ sends a message to $n_a$ and $n_c$ simultaneously tries to send a message to $n_d$, then $n_c$ will find the medium busy although these two signals will never collide because they are destined to opposite directions. Therefore $n_c$ will unnecessarily wait increasing transmission delay in the network.
\subsubsection{How  exposed terminals are detected using classical version protocols}
In \cite{haas02}, several methods are described for hidden and exposed terminal detection. If a node keeps itself informed about identification numbers and locations of two-hop uplink as well as downlink neighbors, then a lot of exposed terminal problems can be resolved.  In this way, $n_a$ will know about locations of $n_b$ and $n_c$; $n_b$ will know locations of $n_a$, $n_c$ and $n_d$. $n_c$ will know about positions of $n_a$, $n_b$ and $n_d$ while $n_d$ will be informed about $n_b$ and $n_c$. In this way, if $n_c$ wants to send a message to $n_d$ and it knows that $n_b$ is sending messages to $n_a$ then $n_c$ will not delay itself because from location information it will identify that these two signals are in opposite direction and won't collide with each other.  But again knowing about two-hop neighbors will require exchanging of HELLO and ACK messages. 

\subsubsection{How  exposed terminals are detected using -HELLO version protocols}
Detecting exposed terminals will require ACK to RREQs. In -HELLO version of protocols, the RREQ message contains information about the location and identifiers of all of its one-hop uplink neighbors. ACK contains all fields of  ACK to HELLO messages along with locations and identification number of its own downlink neighbors. Before sending the first data packet to the specified router in a selected path, the source broadcasts its 1-hop neighbor information. Therefore, for case in figure \ref{Expos ter}, when $n_b$ broadcasts its one-hop neighbor information and information about live communication sessions through $n_b$, then $n_a$ and $n_c$ will know about each other. Along with that, $n_c$ will also be able to identify the directions to which $n_b$ is going to send messages. Unless transmissions of $n_c$ are not in same direction, signals generated by $n_b$ and $n_c$ won't collide. In this way, exposed terminal problems can be solved in -HELLO concept.

\section{SOME -HELLO VERSION PROTOCOLS AS CASE STUDY}
\label{cs}
In the following subsections, we talk about -HELLO versions of AODV, MMBCR, MRPC, MTFR, and MFR namely, -HELLO:AODV, -HELLO:MMBCR, -HELLO:MRPC, -HELLO:MTFR, and -HELLO:MFR.
\subsection{-HELLO:AODV}
\subsubsection{Route Discovery} 
Implementation of -HELLO in reactive routing protocols is convenient because in those protocols, nodes do not have to discover and maintain a route to another node until they
need to communicate. In AODV, nodes use HELLO messages for knowing about local connectivity and there exists one hop-count field in the RREQ message.
Among the various paths through which a RREQ arrives at the destination, the one with the minimum hop count is identified for communication. The minimum hop count value is 1.
The maximum possible hop count value depends upon the total number of nodes in the network. Let it be denoted as $HC$. In -HELLO:AODV, whenever a node $n_j$ receives a RREQ
message from $n_i$, it replies to $n_j$ using an ACK. After receiving the ACK from all downlink neighbors, $n_i$ is capable of constructing its downlink neighbor table which consists of the attributes: \\$<neighbor\_id, neighbor\_location, neighbor\_rad\_range$
and $tmstmp>$.\\ Here $neighbor\_id$, as the name specifies, is the unique identification number of the neighbor; $neighbor\_location$ is an ordered pair that specifies the last known location of the downlink neighbor in terms of latitude and longitude, at timestamp $tmstmp$. With the help of the downlink neighbor table, each node becomes aware of the approximate location of the potential successors and can apply transmission energy optimization during transferring of message packets from one node to another. Hop count field in classical AODV is eliminated in -HELLO:AODV. Each router appends its own node identifier of the RREQ. When the RREQ will arrive at the destination, the destination node will be able to compute the hop count of the path. The hop count is $(\alpha+1)$ where $\alpha$ is the number of router-ids appended to the RREQ. 

Here attributes of the RREQ in -HELLO:AODV are:\\ $< message\_type\_id, source\_id, source\_location, destination\_id, session\_id,\\ number\_of\_data\_packets, initiator\_id, maximum\_hop\_count\_difference, \\router\_ sequence$ and $timestamp >$. \\ Here $message\_type\_id$ is 1 for RREQ messages and 3 for RREP messages in -HELLO:AODV. $source\_id$ and $destination\_id$ specify unique node identifiers of the source and destination nodes. $session\_ id$ is the unique identification number of the communication session between the same pair of source and destination nodes. The trio $<source\_id, destination\_id, session\_id>$ uniquely identify a RREQ. $initiator\_id$ is equal to $source\_id$ if the RREQ message is intended to begin a new communication session or when the link between a source node and its immediate successor has been scrapped and the source wants to discover a new route to the destination. On the other hand, $initiator\_id$ will be an identification number of some router whose link with the corresponding successor (or destination) has been broken. $maximum\_hop\_count\_difference$ field is set to 0 if $initiator\_id=source\_id$ in the RREQ message, that is, maximum hop count for the current path is same as maximum possible hop count in the network and hence their difference is zero; otherwise, $maximum\_hop\_count\_difference$ is $( Z + 1)$ where $Z$ is the number of routers in between the nodes identified by $source\_id$ and $initiator\_id$. For all RREQ messages intending to repair routes, $message\_type\_id$ will be $2$. $message\_type\_id$ is the field that will differentiate between a fresh RREQ and all subsequent route-repair efforts by the source. This information is often helpful for message packet schedulers because route-repair messages are generally given priority over fresh RREQs. The source is definitely excluded. The field number of data packets specifies the number of data packets to send from $source\_id$ to $destination\_id$ in session $session\_id$. \par
Knowing the source location is important for the destination because information about the optimum route selected by the destination, that needs to come back to the source embed in the RREP message. Like classical AODV, whenever a node $n_j$ receives RREQ from $n_i$, it inserts a new entry in the RREQ table where it stores all attributes of the RREQ message except $message\_type\_id$. Also the corresponding predecessor-id (i.e. $n_i$) and predecessor location ($x$ and $y$ coordinates of $n_i$) along with the current timestamp, are stored in the RREQ table. The predecessor information will be required if the link $n_i \longrightarrow n_j$ is present in the optimum path chosen by destination. In that case, $n_j$ shall receive data packets from $n_i$ and send the ACK back to $n_i$. For that, knowing location of $n_i$ will be a prerequisite for $n_j$.

\begin{figure}[htb]
	\centering
	\includegraphics[width=0.65\linewidth]{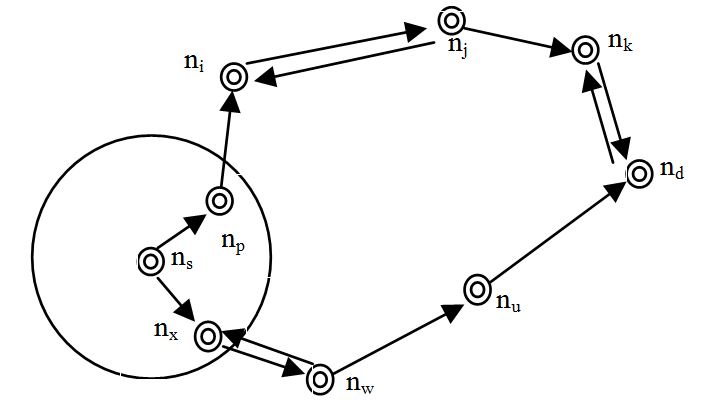}
	\caption{Route establishment from source to destination}
	\label{Route Esta}
\end{figure}

After receiving a RREQ, each node checks whether hop count till that node from the initiator (this can be easily computed from the router's sequence mentioned in RREQ) is less than or equal to the maximum hop count difference mentioned in the RREQ message or not. If a hop count till that node is really less than or equal to the maximum hop count difference mentioned in the RREQ, then only the node process RREQ further for loop detection; otherwise it is readily discarded.
As an example, let us consider figure \ref{Route Esta} where source $n_s$ wants to discover route to a destination $n_d$. Among various paths through which RREQ arrives at the destination, let $ n_s\longrightarrow n_p\longrightarrow n_i\longrightarrow n_j \longrightarrow n_k\longrightarrow n_d $, be a path. Then, $<1, s, (X_s(100), Y_s(100)),d, 3, 5, s, 0, p, 104  >$\\ denote RREQ generated by $n_s$, and  \\$<1, s, (X_s(100), Y_s(100)), d, 3, 5, s, 0, null, 100  >$, \\ $<1, s, (X_s(100), Y_s(100)),d, 3, 5, s, 0, p, i, 108  >$ and \\$<1, s, (X_s(100), Y_s(100)),d, 3, 5, s, 0, p, i,j,k, 115>$\\ forwarded by the routers $n_p, n_i $ and $ n_j$ respectively.
We assume as a typical example that $n_s$ generated the RREQ at timestamp 100 that was processed by nodes $n_p, n_i, n_j, n_k, n_x, n_w, n_u$ at timestamps 104, 108, 110, 115, 104, 107 and 110, in that order.
Similarly, RREQ packets forwarded by routers $n_x, n_w $ and $n_u$ are: $<1, s, (X_s(100), Y_s(100)),d, 3, 5, s, 0, x, 104 >$, \\ $<1, s, (X_s(100), Y_s(100)),d, 3, 5, s, 0, x, w, 107  >$ and \\ $<1, s, (X_s(100), Y_s(100)),d, 3, 5, s, 0, x, w, u, 110  >$ respectively.\\
Assuming that RREQ from $n_s$ arrived at $n_d$ through only the above-mentioned paths, then $n_d$  will choose the route $ n_s \longrightarrow n_x \longrightarrow n_w \longrightarrow n_u \longrightarrow n_d $, because this is having the hop count 4 whereas the earlier path $ n_s \longrightarrow n_p \longrightarrow n_i \longrightarrow n_j \longrightarrow n_k \longrightarrow n_d $, has hop count 5. 

In classical AODV, after processing a RREQ, each router (intending the destination) sets up a reverse path to its predecessor. -HELLO:AODV argues that this is completely unnecessary until and unless the route is really selected for the forwarding of data packets. Classical AODV assumes that most links are bidirectional which is not the case in real life. Therefore, the RREP has to be modeled as another route discovery from destination to source where the sequence of routers in the selected optimum route, will be mentioned. Attributes of RREP are: $<message\_type\_id, destination\_id$ (it specifies the node to which a route was intended to be discovered), $destination\_location, source\_id,  session\_id,\\ initiator\_id, maximum \_hop\_count\_difference, current\_hop\_count,\\ optimum\_router\_sequence$ and $timestamp>.\\ current\_hop\_count$ field is incremented at each router till $HC$ is not reached. The procedure for loop detection is the case of RREP is the same as that in the case of RREQ. No route maintenance is required for RREP because only after the first data packet is sent through the optimum path, routers will know that they are included in the selected path and therefore, need to set a reverse path to the predecessor, that is, the node from which it received the first data packet. Reverse path setup from a node $ n_i$ to $ n_j$ is easy provided the link is bi-directional. Otherwise, directional flooding \cite{benerjee10} is applied to discover a route to $n_i$. That will not incur much cost because the maximum distance between $n_i$ and $n_j$ is radio-range of $n_i$.The format of RREP packets generated by $n_d$ for $n_s$ at timestamp 125 is: $<3, d, (X_d(125), Y_d(125)),x, 3, d, 0, 0, x, w, u, 125 >$.  

\subsubsection{Loop Detection}
After receiving a RREQ a node checks whether hop count till that node is less than or equal to the maximum allowable hop count mentioned in that RREQ. If the condition is satisfied, then the receiver of that RREQ consults its RREQ table, to check whether it has received the same RREQ  earlier. If one such match is found, then a loop is detected and the newly received RREQ is readily dropped. But if a match is found in that table with only difference in $session\_id$ where new $session\_id$ is greater than the previous $session\_id$, then the new entry replaces previous RREQ entry between the same pair of source and destination nodes. But if new $session\_id$ is less than the previous $session\_id$ between the same pair of source and destination nodes, then it denotes that an unnecessary RREQ has arrived. Hence, it can be readily dropped. 

\subsubsection{Route Maintenance}
In classical AODV, if the source node moves during an active session, it has to re-initiate the route discovery procedure to establish a new route to the destination. Similarly, if the destination or some intermediate node moves, a route-break message is sent to the predecessor, because the link from its predecessor to the current router is about to be scrapped. Then the predecessor forwards the route-break message to the source so that the source can initiate a new route discovery session. Route repair becomes necessary only if more data packets are left to be sent to the destination. Periodic HELLO messages are utilized by AODV to detect link failures. On the other hand in  -HELLO:AODV, when a node in a live communication path, is about to leave the radio-range of its predecessor in that path, it sends a proactive link-fail message ($message\_type\_id$ for link-fail is 4) to the predecessor. Attributes of this message are: $<message\_type\_id, source\_id, destination\_id, sender\_id, predecessor\_id$ and $session\_id>$. Here $sender\_id$ is the node which is about to get out of the radio-circle of its predecessor. Receiving the link-fail message, the associated predecessor sends a repair request message to the source of the communication session. Also if the predecessor does not receive ACK of a data packet from its successor within a pre-defined time interval, then it sends a repair-request assuming that the battery of the successor is exhausted.  Attributes of repair-request issued by a router are: $<message\_type\_id$ (5 in case of repair-request) $source\_id, destination\_id,\\ session\_id, link\_break\_timestamp,  initiator\_id, recv\_delay\_source>$. \\All attributes are self-explanatory except $link\_break\_timestamp$ and $recv\_delay\_\\sourc$. $link\_break\_timestamp$ is the timestamp when link breakage was detected by the current router; $recv\_delay\_source$ specifies the time delay that is required by the current node to receive a message from the source. This has been already computed by the current node during the transmission of data packets from the source. All this information greatly helps in reducing multiple simultaneous repairing efforts by different routers to repair the same route. \par
Let us consider the situation when in a route R: $n_s \longrightarrow n_x \longrightarrow n_w \longrightarrow n_u \longrightarrow n_d$, both the links $n_x \longrightarrow n_j$ and $n_w \longrightarrow n_u$ break. Both $n_x$ and $n_w$ will send a repair-request message to $n_s$ as: $< 5, s, d, 3, 130, w, \beta_w>$ and $< 6, s, d, 3, w>$ . Following different cases may occur in that scenario.
\\\textbf{Case-1:}\emph{
	Both links broke at the same time, $n_s$ received repair request from $n_x$}.\\ Here, $n_s$ accords repair-permission to $n_x$.
\\ \textbf{Case-2:}\emph{
	$n_s$ received repair request first from $n_x$ and then from $n_w$}.\\
Here, $n_s$ offers repair permission to $n_x$ only since distance between $n_x$ and $n_s$ is smaller than the same between $n_w$ and $n_s$.
\\ \textbf{Case-3:}\emph{
	$n_s$ permitted route-repair to $n_w$, then received repair request from $n_x$.}\\
In this case, assume that $n_s$ received repair requests of $n_x$ and $n_w$ at timestamps $t_x$ and $t_w$, $t_w < t_x$.  Therefore, repair permission was given to $n_w$ at time $t_w$. Recv-delay-source of $n_x$ and $n_w$ are $\beta_x$ and $\beta_w$, respectively. Therefore, permission granted by $n_s$ is supposed to reach $n_w$ at time $(t_w+\beta_w)$. If $t_x > (t_w+2\beta_w)$, then $n_s$ expects that $n_w$ has received its permission. In that case, $n_s$ keeps  repair-request of $n_x$. Otherwise, permission is granted to $n_x$ too. However, it may happen that, both $n_x$ and $n_w$ gets repair permission from $n_s$ and broadcasts. But that can not cause much harm to the network because $<source\_id, destination\_id, session\_id>$ are same for all those requests and duplicate entries can be easily identified and discarded by routers.
\\ \textbf{Case-4:}\emph{
	None of the routers requesting repair permission could get it from $n_s$.} \\
In this case, $n_s$ won't receive route-reply from $n_d$. Maximum waiting timestamp of $n_s$ is $(t+delRoute+2 \times TTL)$,  where $t$ is the timestamp of generating latest repair permission; $delRoute$ is the time difference between transmitting a data packet from source and getting back its ACK. $TTL$ is  time-to-live of a RREQ.  $(2\times TTL)$ is the maximum time required for a RREQ packet to reach the destination and fetch corresponding RREP back to the source. After timestamp $(t+delRoute+2 \times TTL)\times n_s$, broadcasts RREQ with same $ <source\_id, destination\_id, session\_id>$.

\begin{figure}[htb]
	\centering
	\includegraphics[width=0.7\linewidth]{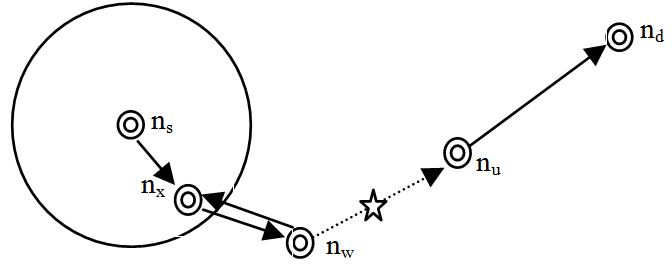}
	\caption{Route establishment from source to destination}
	\label{Route esata2}
\end{figure}

It is expected that the distance between the receiver of repair-permission and the destination is shorter than the same between source and destination. Therefore, a node that receives repair permission, broadcasts an RREQ message to discover a fresh route to the destination, in case more number of packets are to be sent. Due to a comparatively close position with respect to the destination, the number of RREQs produced by the said receiver of repair permission is generally much lesser than the same produced by the source. This is shown in figure \ref{Route esata2}. Link from $n_w$ to $n_u$ is about to be broken because $n_u$ will get out of the radio-circle of $n_w$ very soon. So, $n_u$ sends a link-fail message to $n_w$ and $n_w$ as: $<4, s, d, u, w, 3>$. After receiving the link-fail message, $n_w$ will broadcast RREQ as:$<5, s, d, 3, 130, w, \beta_w>$  to discover a new route to $n_d$. \\
From figure \ref{Route esata2} it can be clearly seen that distance between $n_w$ to $n_d$ is much smaller than distance between $n_s$ and $n_d$. Therefore, if $n_w$ initiate directional route discovery(directional flooding, as the recent location of the destination, is known), then the cost of RREQ packets will be much lesser than if $n_s$ initiates route discovery. If we assume that the link from $n_w$ to $n_u$ broke at timestamp 130 after sending three data packets to the destination and $n_w$ obtained permission for the route-repair from $n_s$ at time 137. Then RREQ generated by $n_w$ will look like $<5, s, d, 3, 130, w, \beta_w>$. Here the number of packets to be transmitted is 2.

\subsubsection{Transmission Power Optimization} 
In AODV, power optimization is performed based on location information of downlink neighbors when they acknowledge HELLO messages of their uplink neighbors. But in -HELLO:AODV, this is not possible because HELLO messages are not periodically sent. Therefore, to implement transmission power optimization in -HELLO:AODV, proactive  ACK is sent from a node to some of its successors; those successors have to be connected to that node through live communication sessions. The interval between two consecutive proactive ACK is the same as the one between two consecutive HELLO messages. Transmission of proactive ACK from $n_j$ to $n_i$  will continue till all communication sessions utilizing link from $n_i$ to $n_j$, complete. Components of this proactive ACK are similar to the HELLO message mentioned in section 3, with only one additional field, namely, minimum-receive-power; the name says it all. Proactive ACK gives $n_i$ information about the most recent location of $n_j$. This is used by $n_i$ while it sends the next data packet to $n_j$, whatever live session it may be.

Let, location of $n_j$ at time $t$ be $(x_j(t), y_j(t))$ where $t$ is timestamp of last proactive ACK from $n_j$ to $n_i$. If minimum received power of $n_j$ be denoted as $minRecv(j)$, then minimum required transmission power $transPower_i(j)$ required by $n_i$ to send a packet to $n_j$ at time $t$, is formulated in \ref{eqn21}. This formulation is as per Frii's transmission equation \cite{kaur17}, \cite{abolhasan04}.
\begin{equation}
	transPower_i(j,t)=minRecv(j) \times dist^2_{ij}(t)/C
	\label{eqn21}
\end{equation}
Where $$
dist_{i,j}(t)=\sqrt{\{X_j(t)-X_i(t)\}^2 + \{Y_j(t)-Y_i(t)\}^2}
$$
Transmission power required by $n_i$ to send a data packet to $n_j$ without power optimization, is denoted as $transNonOpt_i$ and defined in  equation \ref{eqn22}.
\begin{equation}
	transNonOpt_i = minRecv(j) \times \frac{R_i^2}{C}
	\label{eqn22}
\end{equation}
where $R_i$ is radio-range of $n_i$.
Therefore, transmission power $savedPower_i(j,t)$ saved in $n_i$ after optimization based on proactive ACK of $n_j$, at time t, is given by equation \ref{eqn23}.
\begin{equation}
	savedPower_i(j,t)=transNonOpt_i-transPower(j,t)
	\label{eqn23}
\end{equation}  
\subsubsection{Comparing sizes of various messages in AODV and -HELLO:AODV}
During comparison of message sizes in AODV and -HELLO:AODV, first comes the RREQ. Additional attributes in RREQ packet of -HELLO:AODV are: $<initiator\_id, maximum\_hop\_count\_difference$ and $router\_sequence>$. In classical AODV, router sequence was not there because AODV assumed that links are all bi-directional and therefore, maintaining a link to the immediate previous node was sufficient. But in general, link are not bi-directional. So, it is not sufficient to keep track of immediate predecessor because RREP can not always be sent in the reverse link. Hence, in a network consisting of mostly uni-directional links, router-sequence information needs to be maintained. Eliminating router-sequence as an additional RREQ attribute, we are left with $initiator\_id$ and $maximum\_hop\_count\_difference$. If $N$ denotes the total number of nodes in the network, number of bits required to represent initiator-id is $log_2N$. Maximum value of hop count difference is $H$ where $H$ is maximum possible hop count in the network \cite{liu07}. \\Theorem 1 proves that, $
\log_2H=\log_2X+\log_2Y+\log_2\sqrt{X^2+Y^2}-3-\log_2N-\log_2R_{min}$.
So, total number of bits $B\_add\_RREQ$ required to represent the additional attributes of RREQ in -HELLO:AODV, is defined in equation \ref{eqn24}.
\begin{eqnarray}
	B\_add\_RREQ&=&\log_2H+\log_2N \nonumber \\ 
	&=&\log_2X+\log_2Y+\log_2\sqrt{X^2+Y^2}-3-\log_2R_{min}
	\label{eqn24}
\end{eqnarray}
Since, $XY > \sqrt{X^2+Y^2}$ for $X,Y>2$, so, $B\_add\_RREQ<2(\log_2X+\log_2Y)$.
$R_{min}$ is the minimum radio-range among all nodes in the network.
Number of bits required to represent each attribute of HELLO message (it is applicable to classical versions of all the protocols in MANETs), is shown as below:
\begin{enumerate}
	\item message type id($3$ bits)
	\item sender id($\log_2N$ bits)
	\item sender location($\log_2X+\log_2Y$ bits)
	\item radio range ($\log_2R_{max}$ bits)
	\item current time stamp ($\log_2TM$ bits)
\end{enumerate}
$TM$ is the total simulation time and $R_{max}$ is maximum radio-range among all nodes in the network. So, total number of bits $B\_HELLO$ required to represent a HELLO message, appears in equation \ref{eqn25}. 
\begin{equation}
	B\_HELLO=3+\log_2N+\log_2X+\log_2Y+\log_2R_{max}+\log_2TM
	\label{eqn25}
\end{equation}
From equations (25) and (26) we get, $B\_add\_RREQ < 2\times B\_HELLO$\par
It is clear from the above inequality that number of bits required to represent additional attributes in RREQ message is less than two HELLO messages.\\
\textbf{Theorem1}: $\log_2H=\log_2X+\log_2Y+\log_2\sqrt{X^2+Y^2} -3-\log_2N-\log_2R_{min}$\\ 
\textbf{Proof}: Let $P$ and $D$ denote average progress in each hop from source to destination and average distance between a source and destination. Therefore $H$ can be estimated as $H=D/P$. Average one hop progress $P$ is approximated as the maximum distance between a sender and each of the neighbors within its transmission range. Average number of nodes in the circle of radius $R_{min}$, is denoted as $\xi $ and defined in equation \ref{eqn26}. 
\begin{equation}
	\xi=(\frac{N}{(XY)})\pi R^2_{min}
	\label{eqn26}
\end{equation}
The probability of all $\xi$ nodes residing within distance $r$ from center of transmission circle can be formulated as in equation \ref{eqn27}.
\begin{eqnarray}
	F(r)& = & Prob(all \ \xi \; nodes \; residing \;within \;distance \;r) \nonumber \\
	& = & [Prob(a \;node \;reside \; within \; r)]^\xi \nonumber \\
	& = & [\frac{\pi r^2}{\pi R^2_{min}}]^\xi \nonumber \\
	& = & \frac{r^{2\xi}}{R^{2\xi}_{min}}
	\label{eqn27}
\end{eqnarray}
we have assumed independence and randomness node location.
The probability density function (pdf) of progress $r$ from source, is given by equation \ref{eqn28}.
\begin{equation}
	f(r)= \frac{\partial F(r)}{\partial r} = \frac{2\xi r^{2\xi-1}}{R^{2\xi}_{min}}
	\label{eqn28}
\end{equation}
Therefore, average progress is then the expected value of $r$ with respect to pdf $f(r)$, can be calculated as in equation \ref{eqn29}.
\begin{equation}
	I= \int_{0}^{R_{min}} rf(r)\mathrm{d}r=\frac{2\xi R_{min}}{(2\xi +1)}
	\label{eqn29}
\end{equation}
In a network of size $(X \times Y)$, average distance $D$ between source and destination, is approximated as: $$ D\approx \frac{(0 + \sqrt{X^2+Y^2})}{2}$$
Therefore expected number of hops(H) is in equation \ref{eqn30},
\begin{eqnarray}
	H &\approx & \frac{D}{P} \nonumber \\
	&\approx & \frac{\sqrt{X^2+Y^2}}{2I} \nonumber \\
	&\approx & \frac{(2\xi + 1)\sqrt{X^2 + Y^2}}{4\xi R_{min} }
	\label{eqn30}
\end{eqnarray} 
Therefore, 
\begin{multline}
	\log_2H= -\log_22+\log_2X+\log_2Y+\log_2\sqrt{X^2+Y^2}\\-\log_24-\log_2N -\log_2R_{min}
	\label{eqn31}
\end{multline}
Hence by equation \ref{eqn31} proves the theorem.

As far as the link-fail message is concerned, it is completely new in -HELLO:AODV. Number of bits required to represent a link-fail message, is computed as follow:
\begin{enumerate}
	\item message type id($3$ bits)
	\item source id($\log_2N$ bits)
	\item destination id($\log_2N$ bits)
	\item link breakage detector id($\log_2N$ bits)
	\item link broke with node id($\log_2N$ bits)
	\item remaining number of packets ($\log_2PAC$ bits)
\end{enumerate}
Here, $PAC$ is upper limit  of total number of packets that can be transmitted in a session from any source to the destination. Theorem 2 proves that link-fail does not impose any additional byte overhead. \\ 
\textbf{Theorem2}: $Link \; Fail \;does \;not\; require \;any\; additional\; byte.$\\\textbf{}
\textbf{Proof}: If link-fail message is not sent, then data packet has to be sent thrice, that is, two times more than the number of times a data packet is sent in the classical case. Format of a data packet is expressed as follow:
\begin{enumerate}
	\item message type id($2$ bits)
	\item source id($\log_2N$ bits)
	\item destination id($\log_2N$ bits)
	\item sessionr id($\log_2TM$ bits)
	\item packet sequence id ($\log_2PAC$ bits)
\end{enumerate}
Therefore, additional number of bits required to represent a link-fail message compared to two ordinary data packets, is denoted as $B\_add\_Linkfail$ and defined in equation \ref{eqn32}.
\begin{eqnarray}
B\_add\_Linkfail
& =&  3 + 4\log_2N + \log_2PAC \nonumber\\& &- 2( 3 + 2\log_2N + \log_2TM + \log_2PAC ) \nonumber \\&=&  - 3 - \log_2PAC - 2\log_2TM
\label{eqn32}
\end{eqnarray}
Therefore, $B\_add\_Linkfail <0 $ \\
Hence, this proves,  link-fail does not require any additional byte. So, this is an improvement produced by -HELLO:AODV over classical AODV.\par
Repair-request message also exists in classical AODV. The node that discovers link breakage sends a message to the source informing link breakage so that source can initiate route repair. As far as repair-permission message is concerned, it is additional in -HELLO:AODV. Theorem 3 specifies that the additional byte requirement imposed by repair-permission is covered by four HELLO messages.\\ 
\textbf{Theorem 3}: $B\_repair\_permission < (4 \times B\_HELLO).$\\
\textbf{Proof}: Number of bits required by a repair-permission is shown as follow:
\begin{enumerate}
	\item message type id($3$ bits)
	\item source id($\log_2N$ bits)
	\item destination id($\log_2N$ bits)
	\item packet sequence id ($\log_2PAC$ bits)
	\item link breakage detector id($\log_2N$ bits)
\end{enumerate}
So, total number of bits required by a repair-permission  is denoted as $B\_repair\_permission$ and formulated in equation \ref{eqn33}. \\ \\
$B\_repair\_permission = 3 + 3\log_2N + \log_2PAC$ \\   \\
If we assume that $PAC < N$, then,
\begin{equation}
B\_repair\_permission \approx 3 + 3\log_2N
\label{eqn33}
\end{equation}
So, $B\_repair\_permission < (4 \times B\_HELLO$)\\
If $x$  number of HELLO messages are saved, then saved energy $SE$ is given by:\par $SE=x \times avg\_eng\_HELLO(i)$

\subsection{ -HELLO:MMBCR}
\subsubsection{Route Discovery}
In -HELLO embedded version of MMBCR, the source node appends its residual energy information with RREQ message. This field is called minimum-residual-energy. After it is received by the first router, it checks whether its own residual energy is less than that embedded within the RREQ message. If the condition is satisfied, then the router replaces the minimum residual energy in the RREQ message with its own residual energy which becomes the new minimum-residual-energy of the RREQ packet that will be forwarded by the current router. On the other hand, if the residual energy of the current router is greater than or equal to the minimum-residual-energy mentioned in the RREQ it has received, then the current router does not change the minimum-residual-energy of the RREQ packet while forwarding it. Except for minimum-residual-energy, all other fields of the RREQ are similar to -HELLO:AODV. For example, RREQ generated by $n_s$ is as: $<1, s, (X_s(1), Y_s(1)), d, 3, 5, s, e_j(9), r_j, v_j(9), f(s,j), \\s, i, j, 9, 0>$ and network scenario shown in figure \ref{Route Esta}, will look like: $<1, s, (X_s(100), \\Y_s(100)), d, 3, 5, s, 0, null, 100, 4>$ where we have assumed that residual energy of $n_s$ is $4J$. Also assume that residual energy of $n_p$ at timestamp 104 is $2J$ as in: $<1, s, (X_s(100), Y_s(100)), d, 3, 5, s, 0, p, 104, 2>$ and the same of $n_i$ at timestamp 108 is $5J$ shown here:  $<1, s, (X_s(100), Y_s(100)), d, 3, 5, s, 0, p, i, 108, 2>$. \\ Here $n_p$ changed minimum-residual-energy of RREQ sent by $n_s$ from $4J$ to $2J$ because the minimum residual energy of $n_p$ is $2J$ which is less than $4J$. But  $n_i$ did not change the minimum-residual-energy of the RREQ it received from $n_p$ because the residual energy of $n_i$ is $5J$ which is higher than the minimum-residual-energy ($2J$) embedded in RREQ sent by $n_p$ to $n_i$.
RREQs arrive at the destination through multiple paths. All these paths have a minimum-residual-energy. Among them, the path with a maximum of these minimum-residual-energies is selected for communication.\par
Loop Detection, Route Maintenance, and various message sizes of -HELLO:MMBCR are the same as -HELLO:AODV.  

\subsection{-HELLO:MRPC}
\subsubsection{Route Discovery}
In -HELLO  version of MRPC, source node $n_s$ includes an information $f\_Eng(s)$ with the RREQ packet where $f\_Eng(s)$ is defined in equation \ref{eqn34}. 
\begin{equation}
	f\_Eng(s)=\frac{resEng(s)}{unitPktEng(s)}
	\label{eqn34}
\end{equation}
Here, $resEng(s)$ and $unitPktEng(s)$ denote current residual energy of $n_s$ and energy required by $n_s$ to transmit one packet, respectively. After the first router $n_p$ receives that RREQ, it checks whether $f\_Eng(p) < f\_Eng(s)$ or $not$. If as, then $f\_Eng(s)$ is replaced by $f\_Eng(p)$ in the RREQ packet before it is forwarded to the next router. Next router follows a similar procedure. All other fields of the RREQ are same as -HELLO:AODV. For example, RREQ generated by $n_s$ in context of $<1, s, (X_s(1), Y_s(1)), d, 3, 5, s, e_j(9), r_j, v_j(9), f(s,j), s,\\ i,j, 9, 0>$ and network scenario shown in figure \ref{Route Esta}, will look like:$<1, s, (X_s(100),\\Y_s(100)), d, 3, 5, s, 0, null, 100, 1000 >$. Where we have assumed that residual energy of $n_s$ is $4J$. Also assume that residual energy of $n_p$ at timestamp 104 is $2J$ and the same of $n_i$ at timestamp 108 is $5J$. The energy required for transmission of one packet is $4 mJ$ for $n_s, 20 mJ$ for $n_p$ and $10 mJ$ for $n_i$. RREQs forwarded by $n_p$ and $n_i$ are as: $<1, s, (X_s(100), Y_s(100)), d, 3, 5, s, 0, p, 104, 100 >$  and $<1, s, (X_s(100), Y_s(100)), d, 3, 5, s, 0, p,i, 108, 100 >$. Residual packet capacity of $n_s$ is $f\_Eng(s)$ which evaluates to $(4 J / 4 mJ)$ i.e. 1000. The same of $n_p$ is $(2 J / 20 mJ)$ i.e. 100 which is less than the residual packet capacity of $n_s$. Therefore $n_p$ updates the last field of RREQ from 1000 to 100. But  $n_i$ did not change it because $f\_Eng(i)$ is$ (5J / 10 mJ)$ i.e. 500.
RREQs arrive at the destination through multiple paths. All these paths have a residual packet capacity. Among them, the path with maximum $f\_Eng$ is selected for communication.

Loop Detection, Route Maintenance, and various message sizes of -HELLO:MRPC are the same as -HELLO:AODV.

\subsection{-HELLO:MTPR}
\subsubsection{Route Discovery}
In -HELLO:MTPR, source node includes a special transmission power field with RREQ packet which is initially set to null. After the first router $n_p$ receives RREQ packet, it computes minimum transmission power required by $n_s$ to send a message to $n_p$, as in equation (1). This is possible for $n_p$ because $n_p$ knows its own location and minimum receive power requirements. Location of $n_s$, too, is known to $n_p$ from RREQ received from $n_s$. Considering the context of MMBCR, RREQ generated by $n_s$ forwarded by $n_p$ and $n_i$ are as: \\ $< 1, s, (X_s(100), Y_s(100)), d, 3, 5, s, 0, null, 100, null >$, \\ $< 1, s, (X_s(100), Y_s(100)), d, 3, 5, s, 0, p, 104, trans\_power_s(p, 104) >$ and \\ $< 1, s, (X_s(100), Y_s(100)), d, 3, 5, s, 0, p, i, 108, F(s,p,i) >$.\\
Where $F(s,p,i)=min(transPower_s(p,104),transPower_p(i,108))$.\\
RREQs arrive at the destination through multiple paths. All these paths have a minimum transmission power. Among them, the path with the minimum of these minimum transmission powers is selected for communication.

Loop Detection, Route Maintenance, and various message sizes of -HELLO:MRPC are the same as -HELLO:AODV.

\subsection{-HELLO:MFR}
\subsubsection{Route Discovery}
In -HELLO:MFR, source node and each router append their node identifier along with current location so that destination can decide the optimum route based on comparative distances of downlink neighbors of a router, from the router itself. This is projected on the line connecting source and destination of a communication session. Considering the context of MFR, RREQ generated by $n_s$ and forwarded by $n_p$ and $n_i$ are as:$< 1, s, (X_s(100), Y_s(100)), d, 3, 5, s, 0, 100 >$, \\$< 1, s, (X_s(100), Y_s(100)), d, 3, 5, s, 0,p,(X_p(104), Y_p(104)), 104 >$ and \\$< 1, s, (X_s(100), Y_s(100)), d, 3, 5, s, 0,p,(X_p(104), Y_p(104)), i, (X_i(108), Y_i(108)),\\ 108 >$. Other fields are as per AODV counterpart.

\section{DISCUSSION WITH SIMULATION RESULTS}
\label{sm}
\subsection{Simulation Environment}
\begin{table}[htb]
	\caption{Simulation Parameters}
	\begin{center}
		\begin{tabular}{|l|l|}
			\hline 
			Parameter & Specification\\
			\hline 
			Topology Area & 500 x 500(square meter) \\
			
			Traffic type & Constant bit rate (CBR) \\
			
			Packet size	& 512 bytes\\
			
			HELLO packet interval in classical versions & 10 milliseconds \\
			
			Node mobility & 10-30 (meter per seconds) \\
			
			Signal frequency & 2.4 GHz \\
			
			Channel capacity & 2 Mbps \\
			
			Transmission power & 300-600 mW \\
			
			Receiving power & 50-300 mW \\
			
			Mobility model & Random waypoint \\
			
			Radio range & 50 to 100 meter \\
			
			The initial energy of nodes & 5 J to 10 J \\
			
			Pause time	& 1 second \\
			
			Number of nodes	& 20, 40, 60, 70, 80, 90, 100 \\
			\hline
		\end{tabular}
	\end{center}
	\label{simulation}
\end{table}
In simulation experiments, performance analysis of these algorithms is done using network simulation (NS-2) version 2.33. Simulation parameters appear in table 1. -HELLO versions of the protocols AODV, MMBCR, MTPR, MRPC and MFR are compared with their classical versions.
Simulation metrics are energy consumption (in $mJ$), network lifetime ( in $seconds$), average end-to-end delay per session and network throughput (percentage of data packets that could reach their respective destinations)
\begin{figure}[!htb]
	\centering
	\includegraphics[width= 0.7\linewidth]{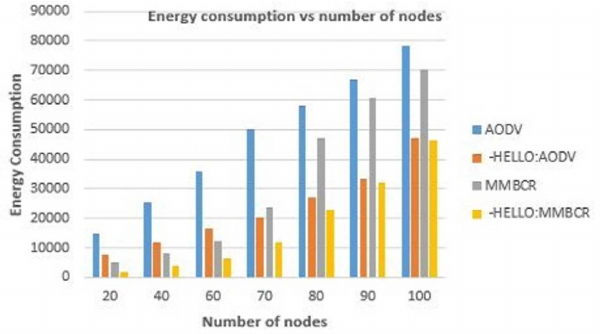}
	\caption{Graphical illustration of energy consumption vs
		number of nodes}
	\label{fig4}
\end{figure}
\begin{figure}[!htb]
	\centering
	\includegraphics[width=0.70\linewidth]{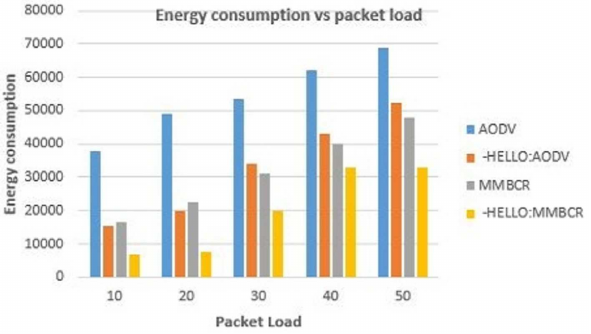}
	\caption{Graphical illustration of energy consumption vs packet load}
	\label{fig8}
\end{figure}
\begin{figure}[!htb]
	\centering
	\includegraphics[width=0.70\linewidth]{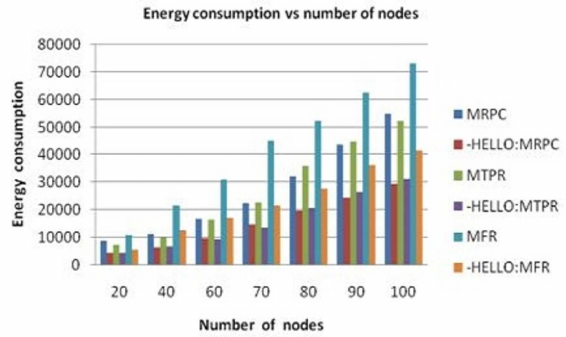}
	\caption{Graphical illustration of energy consumption vs number of nodes}
	\label{fig12}
\end{figure}
\begin{figure}[!htb]
	\centering
	\includegraphics[width=0.70\linewidth]{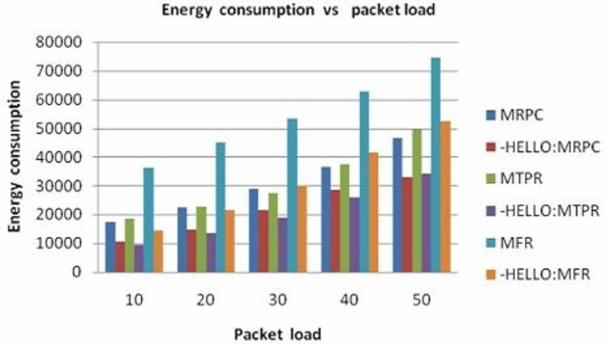}
	\caption{Graphical illustration of energy consumption vs packet load}
	\label{fig16}
\end{figure}
\subsection{Simulation Results}
Simulation graphs appear in the figure \ref{fig4} to figure \ref{fig19}. Explanation these graphs given below with reference to performance metrics namely, energy consumption, network lifetime, end-to-end delay and network throughput. 
\subsection{Energy Consumption}
Compared to classical versions of the protocols (AODV, MMBCR, MRPC, MTPR and MFR) with -HELLO versions, energy consumption in nodes are greatly reduced. It has been shown in this article that by simple alteration of structures in RREQ message, HELLO messages can be
avoided especially in reactive, energy-aware and stability oriented routing protocols. Moreover, route maintenance in -HELLO embedded protocols, is performed in such a manner that it consumes less energy than route maintenance in classical versions of those protocols.
Whenever a link breakage is detected by a router in a live communication path, classical protocols instruct the router to send that information to the associated source of communication, so that the source can re-initiate a route discovery process. -HELLO version protocols emphasize on the fact that distance of current router, that has discovered link breakage, from the destination, is expected to be significantly smaller than the same between source and destination. 
This leaves a deep impact on energy consumption in route re-discovery.
If source initiates route re-discovery, then more RREQ packets will be generated compared to the situation when route rediscovery is initiated by a router than has discovered link breakage. Injection of more RREQ packets means that those packets have to be forwarded by other nodes in the network, increasing energy consumption by nodes.
This is seen by figure \ref{fig4}, figure \ref{fig8}, figure \ref{fig12} and figure \ref{fig16}.  As expected, energy consumption increases with number of nodes and also with increase in packet load (as per figure \ref{fig8} and figure \ref{fig16}).\par
However, energy consumption in AODV is much higher than others because AODV is not concerned with energy of nodes. It selects the path with minimum hop count, as optimal. MFR, although does not directly associate its optimum path selection criteria with residual energies of nodes, but still it tries to minimize pair-wise distance between consecutive routers. So, transmission power required by source and each router is minimum possible in case of MFR. MRPC, MTPR and MMBCR are already energy aware but still eliminating HELLO ensures great improvement. This improvement is 50.17\% for AODV, 41.67\% for MMBCR, 42.46\% for MRPC, 40.48\% for MTPR, 45.25\% for MFR. 
\begin{figure}[!htb]
	\centering
	\includegraphics[width=0.7\linewidth]{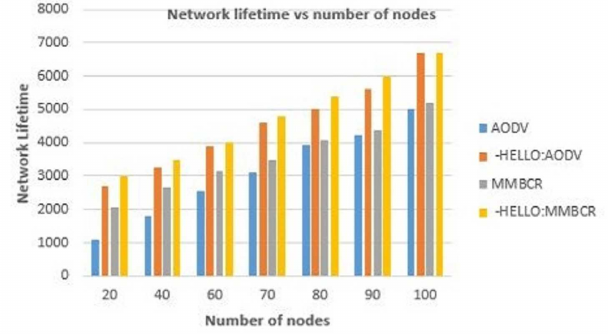}
	\caption{Graphical illustration of network lifetime vs number
		of nodes}
	\label{fig5}
\end{figure}
\begin{figure}[!htb]
	\centering
	\includegraphics[width=0.7\linewidth]{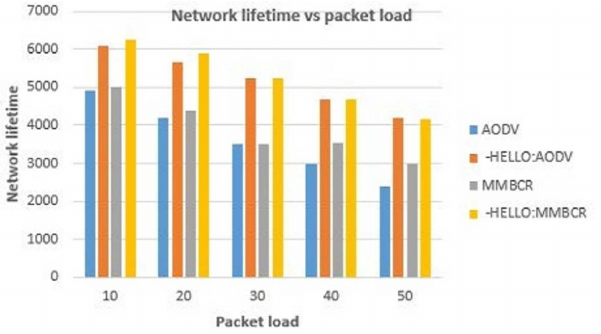}
	\caption{Graphical illustration of network lifetime vs packet load}
	\label{fig9}
\end{figure}
\begin{figure}[!htb]
	\centering
	\includegraphics[width=0.70\linewidth]{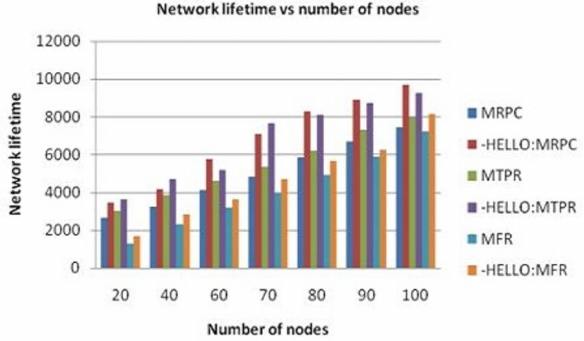}
	\caption{Graphical illustration of network lifetime vs number of nodes}
	\label{fig13}
\end{figure}
\begin{figure}[!htb]
	\centering
	\includegraphics[width=0.70\linewidth]{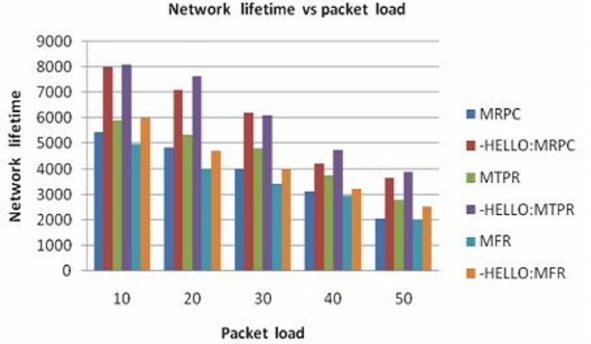}
	\caption{Graphical illustration of network lifetime vs packet load}
	\label{fig17}
\end{figure}
\subsection{Network  Lifetime}
With increase in energy consumption, lifetime of nodes decreases. If a node participating in a live communication dies, then link breakage will be detected by its predecessor and in order to repair the broken link, more RREQ packets are injected into the network. That consumes more energy in nodes resulting in death of more nodes. This is an ominous circle. In schemes lifetime improvements are: 48.48\% for AODV, 39.22\% for MMBCR, 35.56\% for MRPC, 24.57\% for MTPR, 29.64\% for MFR. As seen from the figure \ref{fig5}, figure \ref{fig9}, figure \ref{fig13} and figure \ref{fig17}, network lifetime increases with increase in number of nodes and when number of nodes is fixed and packet load varies, network lifetime reduces.
\begin{figure}[!htb]
	\centering
	\includegraphics[width=0.7\linewidth]{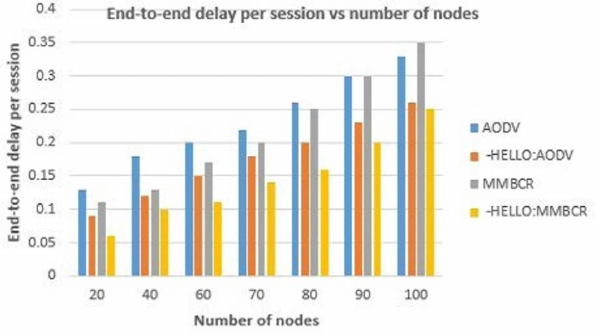}
	\caption{Graphical illustration of end to end delay(in seconds) per session  vs number of nodes}
	\label{fig6}
\end{figure}
\begin{figure}[!htb]
	\centering
	\includegraphics[width=0.7\linewidth]{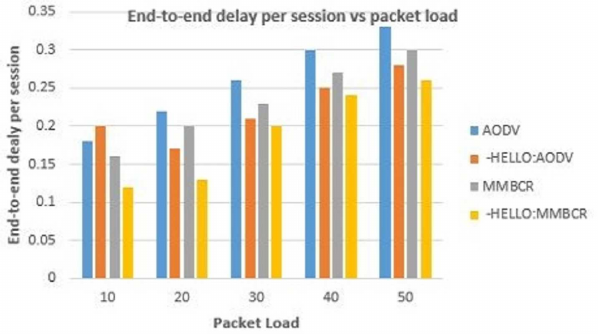}
	\caption{Graphical illustration of end-to-end delay per session vs packet load}
	\label{fig10}
\end{figure}
\begin{figure}[!htb]
	\centering
	\includegraphics[width=0.70\linewidth]{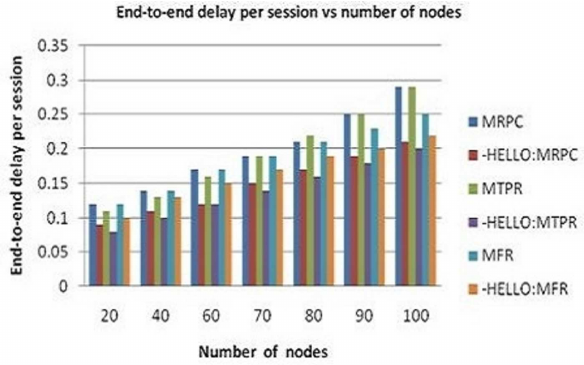}
	\caption {Graphical illustration of end-to-end delay per session vs number of nodes}
	\label{fig14}
\end{figure}
\begin{figure}[!htb]
	\centering
	\includegraphics[width=0.70\linewidth]{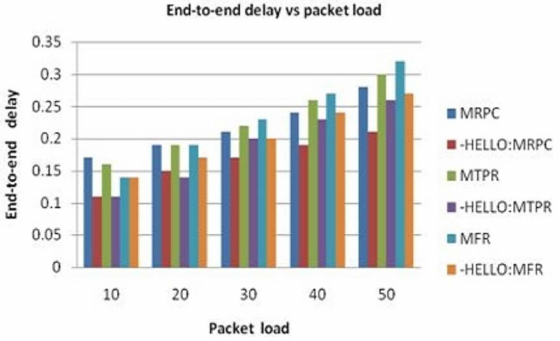}
	\caption{Graphical illustration of network lifetime vs packet load}
	\label{fig18}
\end{figure}
\subsection{End-to-End Delay}
Phenomena like route re-discoveries increase end-to-end delay in a communication session. Reason is that transferring data packets can not start until and unless link breakage is repaired. As mentioned earlier, repairing of link breakage means broadcasting huge number of RREQs and it is a time consuming process. Time duration required for route re-discovery increases end-to-end delay in a communication session. In this scheme delay improvement are: 32.52\% for AODV, 43.45\% for MMBCR, 25.96\% for MRPC, 37.75\% for MTPR, 22.97\% for MFR. The figure \ref{fig6}, figure \ref{fig10}, figure \ref{fig14} and figure \ref{fig18} are referring in this respect.	
\begin{figure}[!htb]
	\centering
	\includegraphics[width=0.7\linewidth]{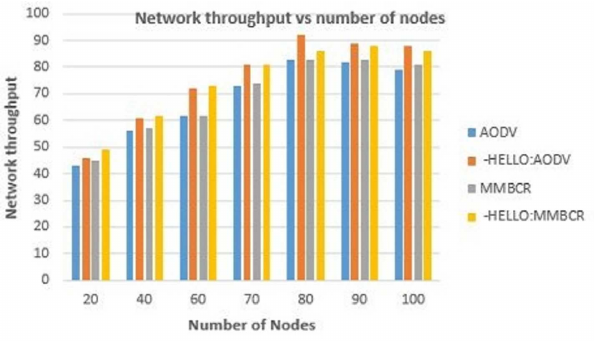}
	\caption{Graphical illustration of network throughput vs
		number of nodes}
	\label{fig7}
\end{figure}
\begin{figure}[!htb]
	\centering
	\includegraphics[width=0.7\linewidth]{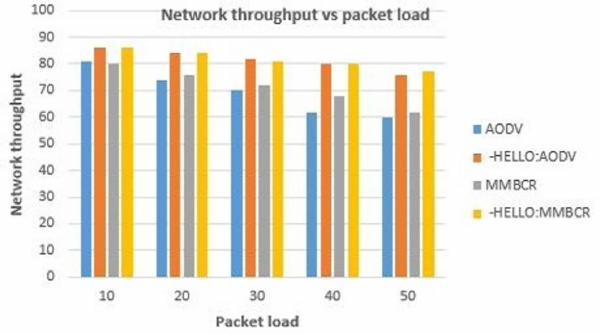}
	\caption{Graphical illustration of network throughput vs packet load}
	\label{fig11}
\end{figure}
\begin{figure}[!htb]
	\centering
	\includegraphics[width=0.70\linewidth]{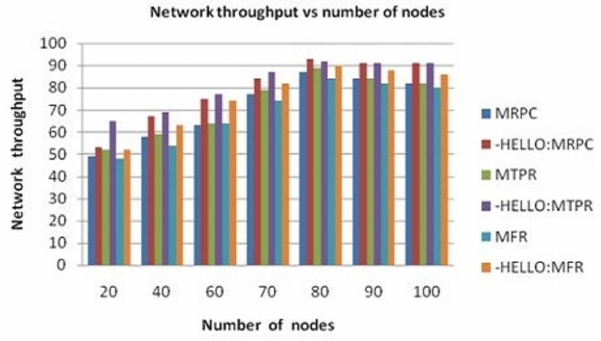}
	\caption{Graphical illustration of network throughput vs number of nodes}
	\label{fig15}
\end{figure}
\begin{figure}[!htb]
	\centering
	\includegraphics[width=0.70\linewidth]{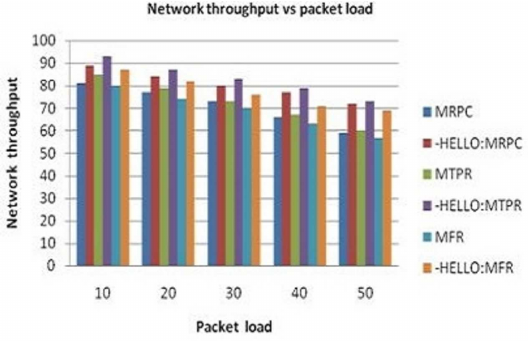}
	\caption{Graphical illustration of network throughput vs packet load}
	\label{fig19}
\end{figure}
\subsection{Network Throughput}
Network throughput is greatly influenced by route re-discovery. An increased amount of injected RREQ packet cause greater number of contention and packet collision in the network. Also network lifetime is reduced and as a result fewer packets can successfully reach their respective destinations. Here throughput improvement are: 11.09\% for AODV, 9.33\% for MMBCR, 8.96\% for MRPC, 9.75\% for MTPR, 8.27\% for MFR. For all the protocols it is seen that initially network throughput increase with increases in number of nodes and later it decreases. Initial improvement is due to better network connectivity whereas after the network becomes dense or highly populated, network throughput starts decreasing. As expected, network throughput decreases with increase in packet load. These findings are evident from the figure \ref{fig7}, figure \ref{fig11}, figure \ref{fig15} and figure \ref{fig19}.

\section{CONCLUSION}
\label{con}
From the perspective of mobile ad-hoc communications, it is extremely important to save battery power as much as possible. That will lead to an increase in the lifetime of nodes ensuring thereby prolonged opportunity to forward packets (both data and control) of others. This -HELLO devoid version protocols reduce energy consumption by reducing or eliminating the HELLO messages.  It also reduces the link breakage phenomenon as most of the link breakage occurs due to exhausted batteries' power of nodes. Therefore, the number of RREQ packets injected into the network for repairing those routes, also reduce. This also leads to a decrease in end-to-end delay, increase throughput. The simulation analysis showing good performance improvements. Extensive analysis and testing in a real testbed IoT enabled environment could be one of the future scopes of the scheme. 

\bibliography{mybib}
\bibliographystyle{elsarticle-num}
\end{document}